# A Simple and Generic Paradigm for Creating Complex Networks Using the Strategy of Vertex Selecting-and-Pairing

Shuangyan Wang, Gang Mei*

School of Engineering and Technology, China University of Geosciences (Beijing)

Email: sy.wang@cugb.edu.cn; gang.mei@cugb.edu.cn

**Abstract**

In many networks of scientific interest we know that the link between any pair of vertices conforms to a specific probability, such as the link probability in the Barabási-Albert scale-free networks. Here we demonstrate how the distributions of link probabilities can be utilized to generate various complex networks simply and effectively. We focus in particular on the problem of complex network generation and develop a straightforward paradigm by using the strategy of vertex selecting-and-pairing to create complex networks more generic than other relevant approaches. Crucially, our paradigm is capable of generating various complex networks with varied degree distributions by using different probabilities for selecting vertices, while in contrast other relevant approaches can only be used to generate a specific type of complex networks. We demonstrate our paradigm on four synthetic Barabási-Albert scale-free networks, four synthetic Watts-Strogatz small-world networks, and on a real email network with known degree distributions.

## 1 Introduction

Complex networks can be exploited to address complex problems in many research fields, such as the epidemiology, ecology, communication, human behavior, biologic, economics, logistic system, internet of things, transportation system [1-4]. The scale-free network and the small-world network are the two classical complex networks. Scale-free networks have been shown to exist in a variety of real-world systems [2], including the social networks, many kinds of computer networks, some financial networks [5, 6], protein-protein interaction networks, and semantic networks [7]. Small-world properties also have been found in many real-world phenomena [8], such as the websites with navigation menus, food webs [9], social influential networks, cultural networks [10], and co-occurrence networks [11].

In the network science, one of the most critical issues is to generate various complex networks. And several methods have been proposed to build complex networks. For example, the Barabási Albert Model (BA) [2] was proposed for generating scale-free network. And there are two generic mechanisms in BA model: (1) networks expand continuously by addition of new vertices, and (2) new vertices attach preferentially to sites that are already well connected. It is different from the linear preferential attachment of BA model, P.L. Krapivsky et al. [12] proposed a growing random network model with a non-linear preferential attachment according to a rate equation approach. The network is built by adding sites that link to earlier sites with a probability $A_k$ which depends on the number of preexisting links $k$ to that site. And scale-free networks can be generated by tuning the parameters of the rate equation. Moreover, Dangalchev et al. [13] proposed a 2-L model by adding a second-order preferential attachment. The essential idea of the 2-L model is that a vertex was connected with neighbors according to the degrees of neighbors and the degrees of the connected neighbors of its neighbors. And Pachon et al. [14] proposed a Uniform-Preferential-Attachment model (UPA model) for generating scale-free networks according to two attachment rules: a preferential attachment mechanism (with probability 1-p) that stresses the rich get richer system, and a uniform choice (with probability p) for the most recent vertices. In the field of biologic, many approaches have been presented to generate scale-free networks by using the duplication and divergence initialization [15, 16].

In addition to the models for generating scale-free networks, the Watts–Strogatz model (WS model) and the Newman-Watts model (NW model) are two classical models for generating small-world networks. The WS model [17] added a randomly rewired-edge mechanism based on a regular ring lattice, and rewiring was done by replacing with where *k* is chosen uniformly at random from all possible nodes while avoiding self-loops and link duplication. The NW model [18] added a randomly added-edge mechanism based on a regular ring lattice while avoiding self-loops and link duplication. Essentially, the NW model is the same as the WS model.

From the above-described models for generating scale-free networks and small-world networks, a generic principle could be achieved, i.e., to link vertices according to a random selecting probability. The random probability is utilized to select the preferential linked neighboring vertices. And different probabilities for selecting neighbors could lead to different complex networks.

In this paper, we propose a generic and straightforward paradigm for creating complex networks by using the strategy of vertex selecting-and-pairing on the basis of exploiting different distributions of selecting probabilities. To examine the performance of the proposed paradigm, we create the classical BA scale-free networks, the WS small-world networks, and a scale-free network which is similar to a real email network by exploiting the proposed paradigm.

The rest of this paper is organized as follows: Section 2 gives a detailed introduction to the proposed generic paradigm. Section 3 and Section 4 present and discuss the results of our designed tests, respectively. Section 5 draws several conclusions.

## 2 Methods

The essential idea behind the proposed paradigm for generating complex networks is to select two vertices of an edge according to probability distributions. Moreover, different combinations of the probability distributions can lead to different eventual degree distributions in generated networks.

There are three steps in the paradigm, including (1) a procedure for avoiding isolated vertices, (2) a procedure for selecting-and-pairing vertices, and (3) a cleaning procedure for removing self-loops and link duplications.

Moreover, there are three input parameters of the proposed paradigm, including (1) the number of vertices, (2) the number of edges, and (3) the expected degree distribution.

### 2.1 Procedure for Avoiding Isolated Vertices

In this subsection, we propose two schemes for avoiding isolated vertices, including Scheme A and Scheme B.

### 2.1.1 Scheme A for Avoiding Isolated Vertices

To avoid isolated vertices in the generated networks, we conduct a simple preprocessing before creating complex networks. The essential idea is to link each vertex only once to any vertex of the rest vertices. For example, the vertex indexed with 0 will be linked to the vertex indexed with 1; and this is the same for the two vertices indexed with 2 and 3; see **Figure 1(a)**. The degrees of all vertices are 1 after this preprocessing.

When the number of vertices, i.e., *m*, is even, then ($m$ / 2) edges could be created. When m is odd, then the last vertex will be linked to the first vertex, and thus ($m$ / 2 + 1) edges could be obtained. And in this case, all the vertices have the degree of 1 except the first vertex has the degree of 2.

### 2.1.2 Scheme B for Avoiding Isolated Vertices

The essential idea behind the Scheme B is to orderly select the first vertices of all edges and to randomly select the corresponding second vertices of all edges based on probabilities, see **Figure 1(b)**. Moreover, the vertices which have been linked will be marked, and the marked vertices cannot be selected as the first vertex of any edge again.

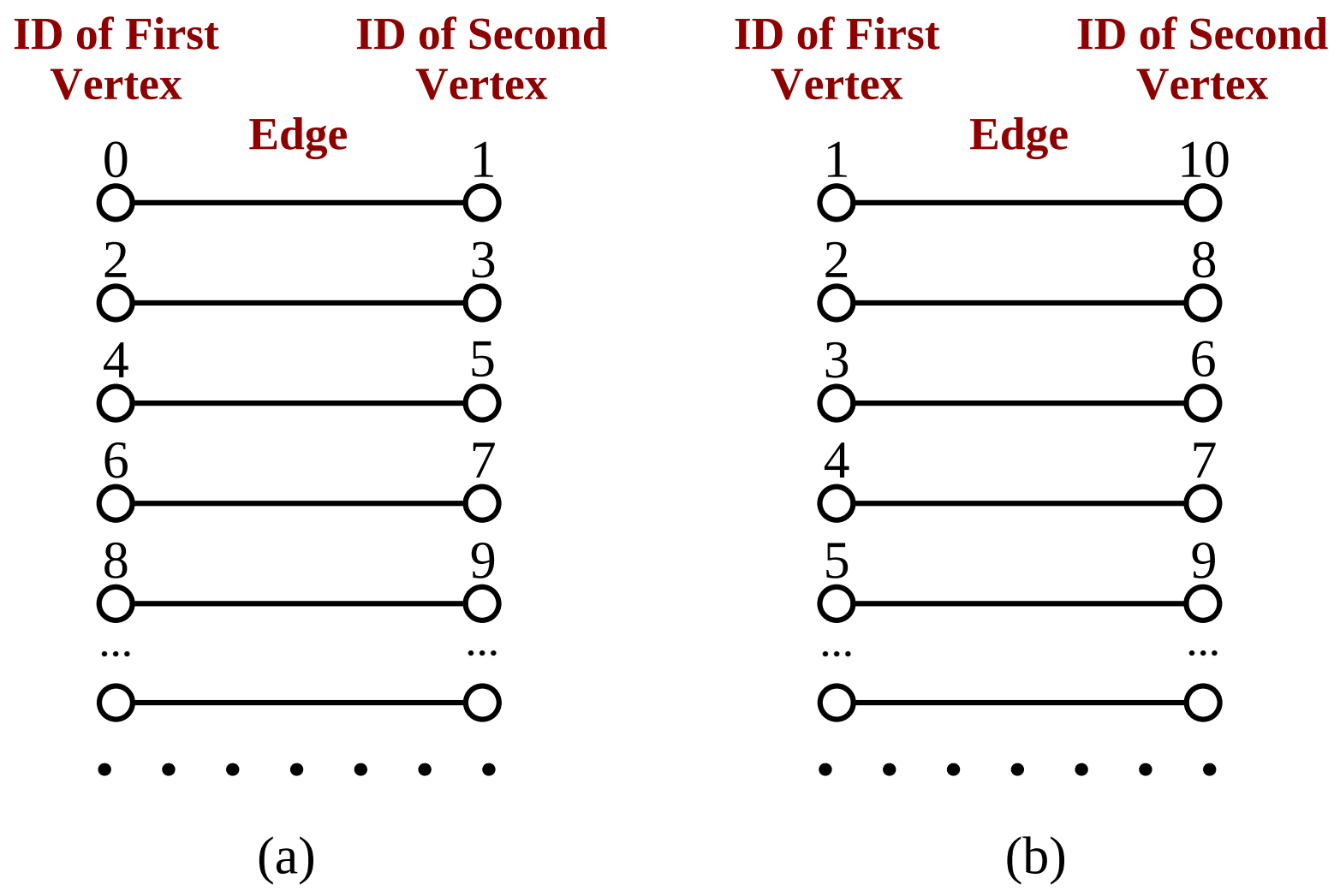

**Figure 1**. Illustrations of the preprocessing for avoiding isolated vertices. (a) Scheme A for avoiding isolated vertices; (b) Scheme B for avoiding isolated vertices.

## 2.2 Procedure for Selecting-and-Pairing Vertices

After the above preprocessing procedure, we further create the desired complex network by selecting-and-pairing vertices. The essential idea is to select two vertices which linked by an edge according to different distributions of selecting probabilities. That means, the probabilities for selecting two vertices which linked by an edge conform to a specific statistic distribution. For example, when the probabilities for selecting two vertices which are linked by an edge can all conform to the Uniform distributions, then the small-world network will be generated. When the probabilities for selecting two vertices which are linked by an edge can all conform to the Gaussian distributions, then the scale-free network will be generated.

Note that the distributions of the selecting probabilities for selecting two vertices can be different. For example, one of the two vertices is selected according to selecting probabilities which conform to a Uniform distribution and the other vertex is selected according to a Gaussian distribution. And the different combinations of the distributions of selecting probabilities also can give rise to different complex networks.

Suppose there are *m* vertices and *n* edges in the expected complex network, and there are *s* edges which remain after the first procedure. The procedure for selecting-and-pairing vertices is composed of: (1) randomly selecting two vertices which linked by an edge from the *m* vertices, respectively, and the random probabilities for selecting two vertices conform to two statistic distributions which can be different or the same, see **Figure 2**; (2) looping over all remain edges, and pairing two vertices for each of the *s* edges according to Step (1).

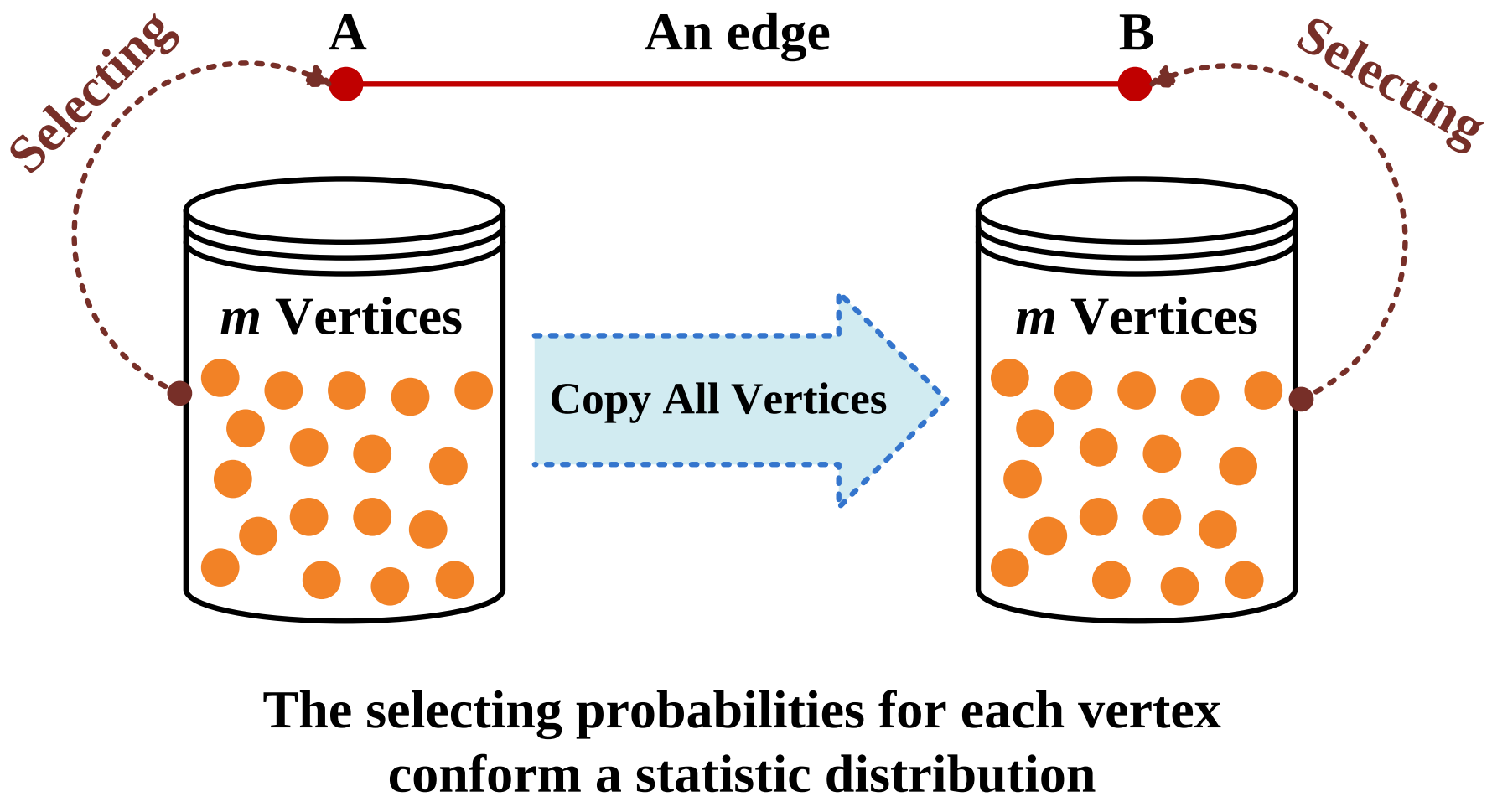

**Figure 2**. Illustration of the procedure for selecting-and-pairing vertices

### 2.3 Procedure for Removing Self-loops and Link Duplications

After the above two procedures, it needs to further remove self-loops and link duplications in the expected network. The specific procedure is composed of:

(1) Copying all edges with the opposite direction; see Figure 3(a). For instance, the copied vertex A is the original vertex B and the copied vertex B is the original vertex A.

(2) Sorting all edges first according to the IDs of the vertex A and then according to the IDs of the vertex B in ascending order.

(3) Comparing each pair of arbitrarily adjacent edges and removing one of the two edges which have the same two vertices. For instance, in Figure 3(b), the left part of Figure 3(b) illustrates the comparison of two vertices of arbitrarily adjacent two edges, the second edge and the third edge have the same two vertices, i.e., the vertices $A_2$ and $B_2$. The right part of Figure 3(b) illustrates the results of the comparison, and one of the original second and the third edges remains, and one of them is removed.

(4) Removing those edges who have the same vertices (i.e., the vertex A = vertex B) and those edges whose ID of vertex A is larger than the ID of vertex B (i.e., the vertex A > vertex B).

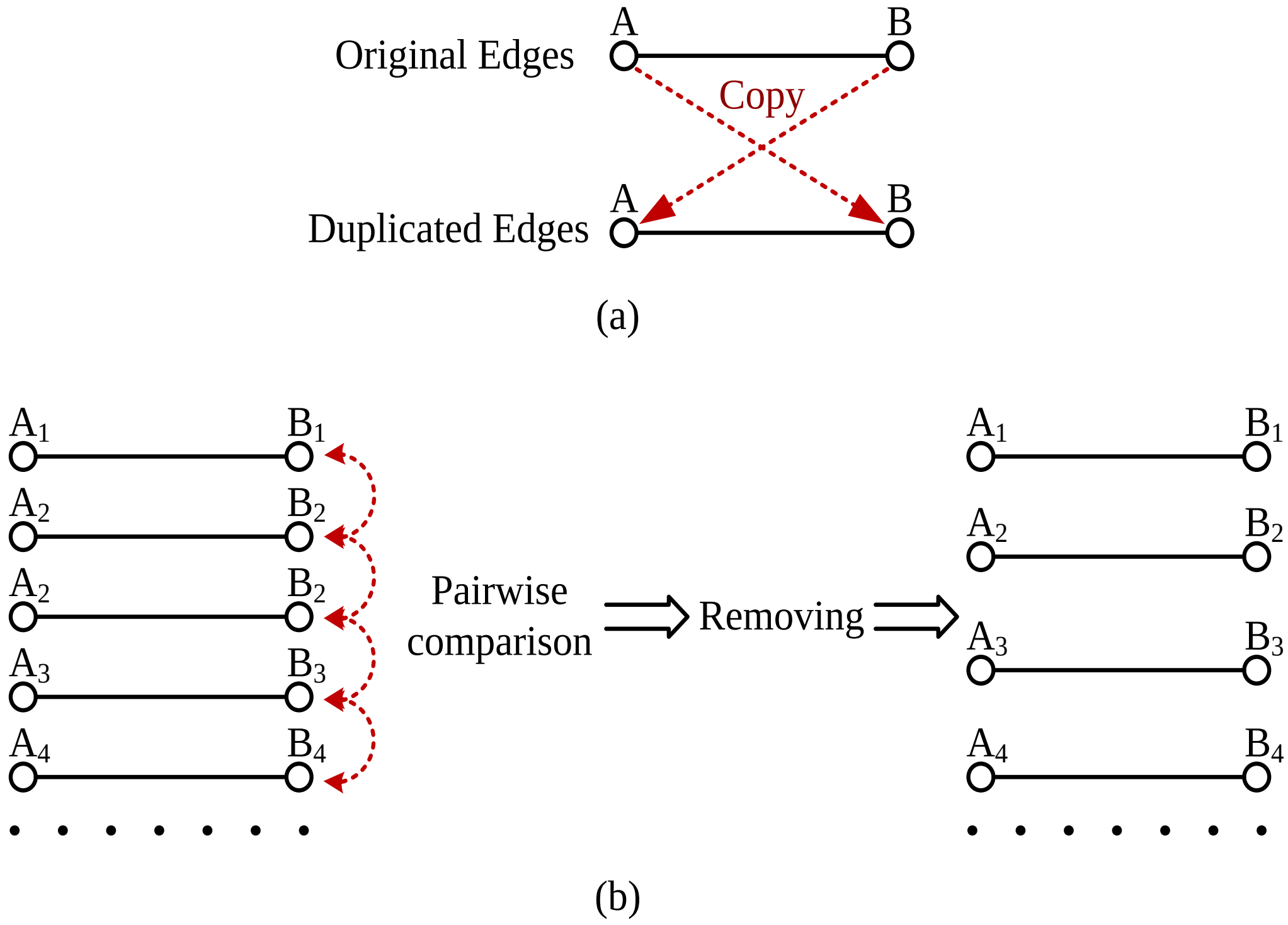

**Figure 3**. Illustrations of the procedure for removing self-loops and link duplications. (a) Copying all edges with opposite direction; (b) Comparing all pairs of adjacent edges to remove link duplications.

## 3 Results

In this section, we give an introduction to the experimental design and test data. And we present three groups of experimental results, including (1) the performance of creating BA scale-free networks, (2) the performance of creating WS small-world networks, and (3) the performance of creating a real email network.

### 3.1 Experimental Design and Test Data

To create the scale-free network and small-world network, we select the Uniform distribution and the Gaussian distribution for the selecting probabilities. We could use several combinations of distributions to create the scale-free networks and small-world networks; see **Table 1**.

For the Scheme B of the first procedure, there are two distributions can be configured, i.e., the second vertex of each edge can be selected according to the probabilities conforming to the Uniform or Gaussian distribution. And in the second procedure, there are three combinations of two selecting probability distributions for two vertices which linked by each edge, respectively, i.e., the probabilities for selecting two vertices which are linked by each edge (1) all conform to the Uniform distributions, or (2) all conform to the Gaussian distributions, or (3) one of them conforms to the Uniform distribution and one of them conforms to the Gaussian distribution. Therefore, there are 9 combinations for creating scale-free networks and small-world networks, i.e., the combination P0D1,

P0D2, P0D3, P1D1, P1D2, P1D3, P2D1, P2D2, and P2D3; see **Table 4**.

Moreover, there is an input parameter of the Gaussian distribution, i.e., the variance $\sigma$. The $\sigma$ can affect the generated network structure, and we configure the range of $\sigma$ being from 0.1 to 1.0, and the uniform interval is 0.1.

There are 9 benchmark networks, i.e., (1) four BA scale-free networks with 4 sizes, (2) four WS small-world networks with 4 sizes, and (3) an email scale-free network [19, 20] with 1005 vertices and 25571 edges which is from a large European research institution. More details on the 9 benchmark networks are listed in **Table 2**. The four BA networks and four WS networks are generated by the software Anylogic [21] which is a multi-agent modeling platform.

**Table 1.** Combinations for creating the scale-free networks and small-world networks

<table>
<tr><th>Procedures of the Proposed Paradigm</th><th colspan="2">Distribution of Selecting Probabilities</th><th>Label</th></tr>
<tr><td rowspan="3">The First Procedure</td><td colspan="2">Scheme A</td><td>P0</td></tr>
<tr><td rowspan="2">Scheme B</td><td>Uniform</td><td>P1</td></tr>
<tr><td>Gaussian</td><td>P2</td></tr>
<tr><td rowspan="6">The Second Procedure</td><td>Vertex A</td><td>Uniform</td><td rowspan="2">D1</td></tr>
<tr><td>Vertex B</td><td>Uniform</td></tr>
<tr><td>Vertex A</td><td>Gaussian</td><td rowspan="2">D2</td></tr>
<tr><td>Vertex B</td><td>Gaussian</td></tr>
<tr><td>Vertex A</td><td>Uniform</td><td rowspan="2">D3</td></tr>
<tr><td>Vertex B</td><td>Gaussian</td></tr>
</table>

**Table 2.** Details of the adopted nine benchmark networks

<table>
<tr><th>Benchmark Networks</th><th>Number of Vertices</th><th>Number of Edges</th><th colspan="2">Input Parameters</th></tr>
<tr><td rowspan="4">BA Networks</td><td>5000</td><td>4999</td><td rowspan="4">Number of Hub</td><td rowspan="4">1</td></tr>
<tr><td>10000</td><td>9999</td></tr>
<tr><td>20000</td><td>19999</td></tr>
<tr><td>40000</td><td>39999</td></tr>
<tr><td rowspan="4">WS Networks</td><td>5000</td><td>13443</td><td rowspan="2">Average Links</td><td rowspan="2">3</td></tr>
<tr><td>10000</td><td>27023</td></tr>
<tr><td>20000</td><td>53946</td><td rowspan="2">Linked Probability</td><td rowspan="2">0.8</td></tr>
<tr><td>40000</td><td>108275</td></tr>
<tr><td>Real Email Network</td><td>1005</td><td>25571</td><td colspan="2">/</td></tr>
</table>

### 3.2 Performance of Creating BA Scale-free Networks

There are 4 benchmark networks employed for evaluating the performance of creating BA scale-free networks; see **Table 2**.

The degree distribution of BA scale-free networks is the Power-low distribution; see the probability density function in **Equation (1)** [22]. The range of $b$ is from 2 to 3, and the detailed values of the parameters $a$ and $b$ in the 4 benchmark BA scale-free networks are listed in **Table 3**.

$$P(k) = ak^{-b}$$

Equation (1)

(P($k$) expresses the probability that randomly selecting a vertex with $k$ degree from all vertices.)

**Table 3**. Values of the parameters $a$ and $b$ in the 4 benchmark BA scale-free networks

| Network Size | 5000 | | 10000 | | 20000 | | 40000 | |
|---|---|---|---|---|---|---|---|---|
| | a | b | a | b | a | b | a | b |
| **Benchmark Networks** | 0.666 | -2.060 | 0.666 | -2.078 | 0.669 | -2.082 | 0.670 | -2.097 |

When employing the combinations P0D2, P0D3, P1D1, P1D2, P1D3, P2D1, P2D2, and P2D3, the networks which have the Power-low degree distribution will be generated. It has been observed that all of the above 8 combinations involving the probabilities conforming to Gaussian distribution for selecting vertices.

Moreover, we find that the value of $b$ always decreases with the increase of the $\sigma$ in the Gaussian distribution. It is possibly because that the discrete of the Gaussian distribution increases and the kurtosis of the Gaussian distribution decreases with the increase of the variance $\sigma$. In this case, the differences between the selecting probabilities of vertices increase, and the power exponent $b$ decreases.

However, the changes of $b$ are varied in different combinations. For example, the decreasing rate of $b$ in the combination P0D2 is more than that in the combination P0D3, see **Table 4**. We perform the differences of the changes of $b$ based on the value of $\sigma$ while $b$ = 2. And the "/" expresses that there is no $b$ = 2 in the range of the $\sigma$. And we rank the eventual Power-low distributions according to the value of $\sigma$ while $b$ = 2, i.e., the Power-low distribution with small $b$ (while there is no $b$ = 2), middle $b$ ($\sigma \leq 0.5$), large $b$ ($\sigma > 0.5$). More details and test data are listed in the **Appendix**.

The above behavior is probably due to (1) the different combinations of distributions for selecting vertices in the second procedure and (2) the different combinations of the first and the second procedures.

For example, in **Table 1** and **Table 4**, (1) in the procedure D1, double Uniform distributions can lead to a Poisson distribution; (2) in the procedure D2, double Gaussian distributions can lead to a Power-low distribution with the large $b$; (3) in the procedure D3, Uniform and Gaussian distributions can lead to a Power-low distribution with the small $b$; (4) in the combination of P1 and D2, Uniform and the Power-low with the large $b$ can lead to a Power-low with the small $b$; (5) in the combination of P2 and D2, Gaussian and the Power-low with the large $b$ can lead to a Power-low with a large $b$; and (6) in the combination of P2 and D1, double Gaussian can lead to a Power-low distribution with a middle $b$.

**Table 4**. The degree distributions when using different combinations

| NO. | Combination | Procedure | Degree Distribution In Procedure | Degree Distribution In Combination | $\sigma$ (b = 2) |
|---|---|---|---|---|---|
| **1** | P0D1 | P0 | No Distribution | Poisson | N/A |
| | | D1 | Gaussian | | |
| **2** | P0D2 | P0 | No Distribution | Power-low | $\sigma > 0.5$ |
| | | D2 | Power-low | | |
| **3** | P0D3 | P0 | No Distribution | Power-low | N/A |
| | | D3 | Power-low | | |
| **4** | P1D1 | P1 | Uniform | Power-low | N/A |
| | | D1 | Gaussian | | |
| **5** | P1D2 | P1 | Uniform | Power-low | N/A |
| | | D2 | Power-low | | |
| **6** | P1D3 | P1 | Uniform | Power-low | N/A |
| | | D3 | Power-low | | |
| **7** | P2D1 | P2 | Gaussian | Power-low | $\sigma \leq 0.5$ |
| | | D1 | Gaussian | | |
| **8** | P2D2 | P2 | Gaussian | Power-low | $\sigma > 0.5$ |
| | | D2 | Power-low | | |
| **9** | P2D3 | P2 | Gaussian | Power-low | $\sigma \leq 0.5$ |
| | | D3 | Power-low | | |

Based on the analysis of the changes of $b$ in different combinations, we can generate the BA scale-free networks which have the almost the same parameters $a$ and $b$ as those in the benchmark BA scale-free networks by using 4 combinations with different $\sigma$, i.e., the combination P0D2, P2D1, P2D2, and P2D3 with $\sigma = 0.95$, $\sigma = 0.32$, $\sigma = 1.1$, and $\sigma = 0.55$, respectively. The fitted curves of degree distributions in the generated and benchmark BA networks are illustrated in **Figure 4**.

For other combinations, i.e., the combinations P0D2, P1D2, and P1D3, the benchmark BA scale-free networks cannot be generated by using those combinations. It is caused by the changes of $b$ with the increase of $\sigma$ in those combinations. In those combinations, the maximum $b$ while $\sigma = 0.1$ is smaller than 2, and the further decreased $\sigma$ cannot lead to the further increased $b$.

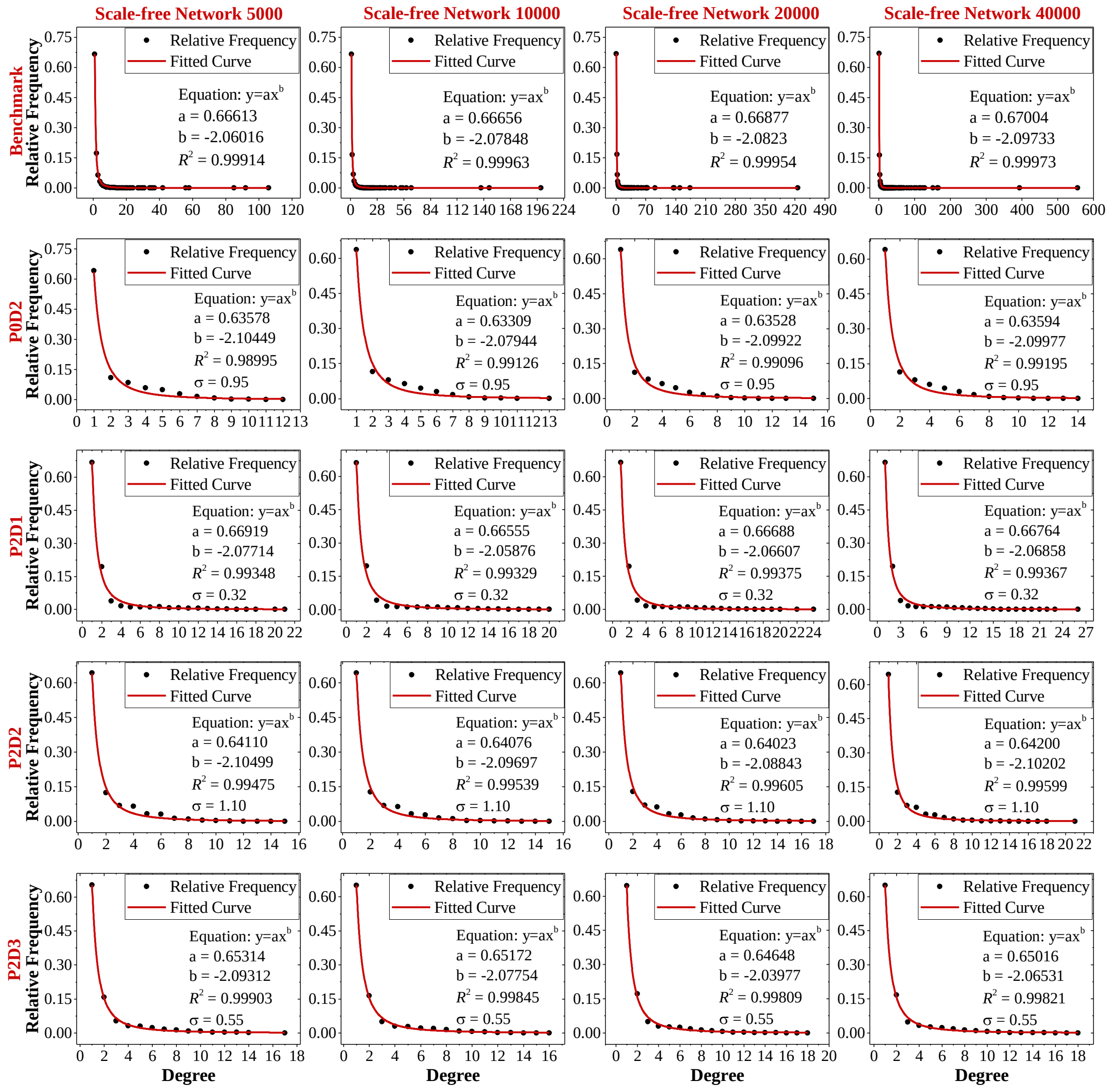

**Figure 4**. Fitted curves of degree distributions in the generated and benchmark BA networks

### 3.3 Performance of Creating WS Small-world Networks

There are four benchmark networks employed for evaluating the performance of creating WS small-world networks; see **Table 2**.

The degree distribution of WS small-world networks is the Poisson distribution, see the probability density function in **Equation (2)** [22]. The values of $r$ in the four benchmark small-world networks with 5000, 10000, 20000, and 40000 vertices are 5.509, 5.544, 5.529, and 5.415, respectively.

$$\mathrm{P}(k) = \frac{r^k e^{-r}}{k!}$$

Equation (2)

($\mathrm{P}(k)$ expresses the probability that randomly selecting a vertex with $k$ degree from all vertices.)

When employing the combination P0D1, the networks which have the Poisson degree distribution will be generated. The fitted curves of degree distributions in the generated and benchmark WS networks are illustrated in **Figure 5**.

It has been observed that the values of $r$ in the generated networks are close to the values of $r$ in the benchmark networks; and for the Poisson fitted curves, the Goodness of Fit in the generated networks is better than that in the benchmark networks.

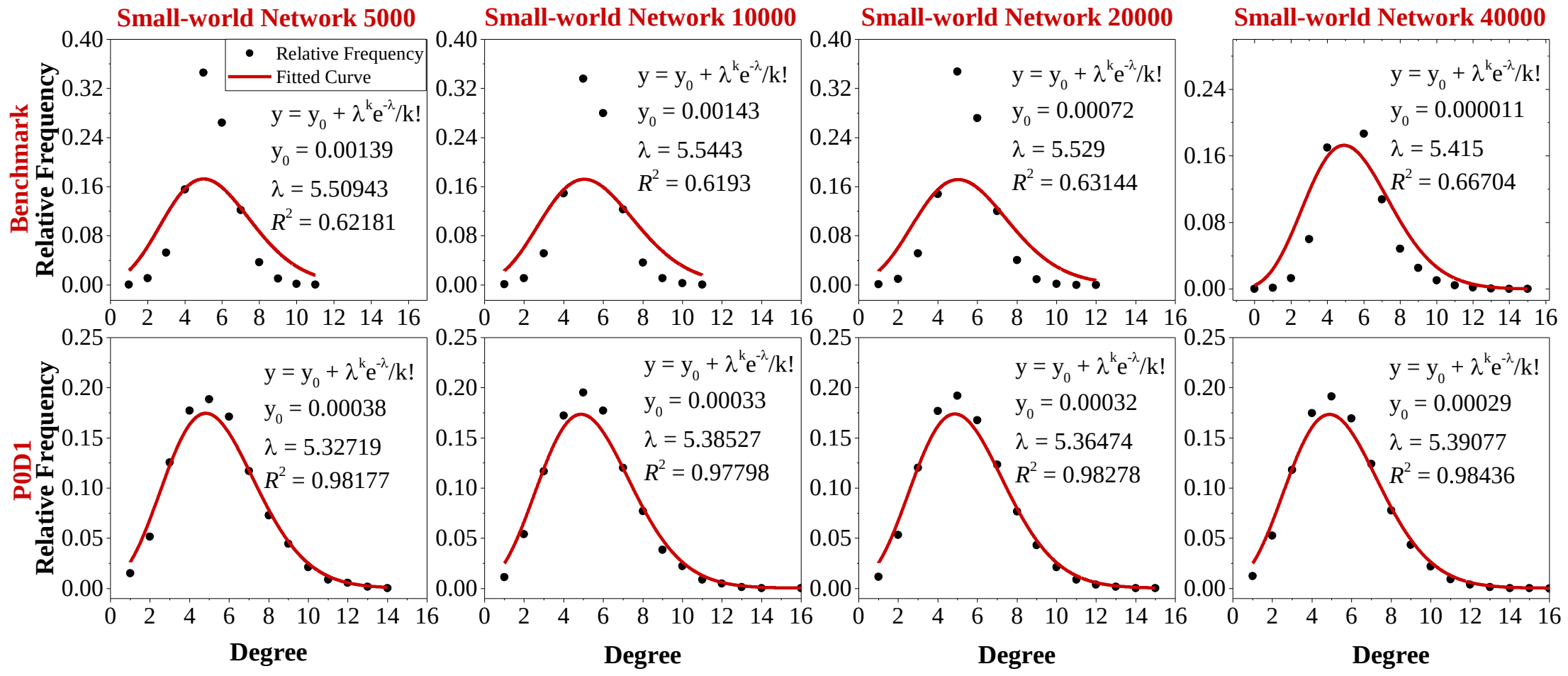

**Figure 5**. Fitted curves of degree distributions in generated and benchmark WS networks

### 3.4 Performance of Creating a Real Email Network

There is only one benchmark network employed for evaluating the performance of creating a real email network; see **Table 2**.

The degree distribution of the real email network is the Power-law distribution, see the probability density function in **Equation (1)** [22]. The parameters $a$ and $b$ in **Equation (1)** are 0.086 and 0.697, respectively.

With the use of input parameters of the real email network listed in **Table 2**, the expected networks which have the Power-low degree distribution can be generated by using the combinations P0D2, P1D2, and P2D2 with $\sigma$ = 1.4, $\sigma$ = 1.45, and $\sigma$ = 1.4, respectively, see **Figure 6**.

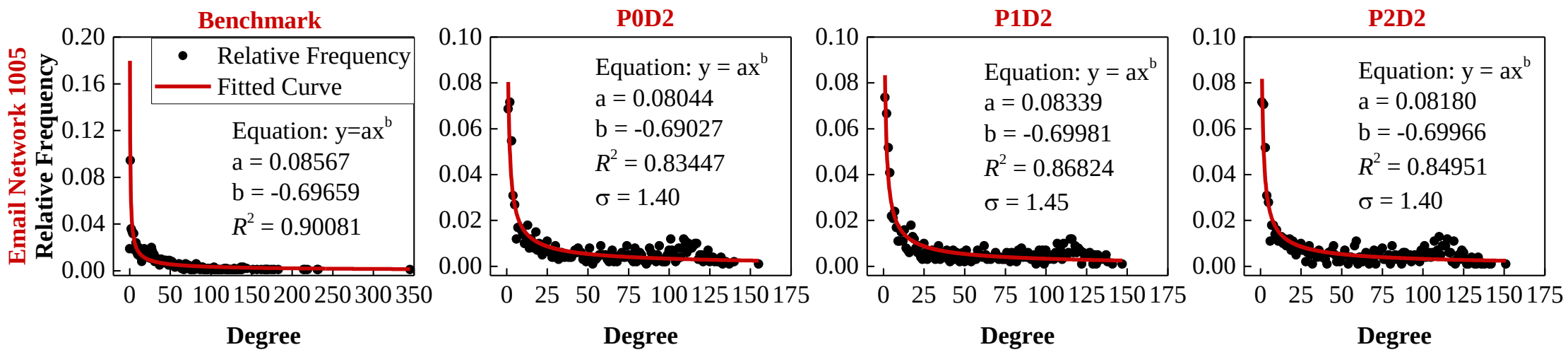

**Figure 6**. Fitted curves of degree distributions in the generated and benchmark email networks

We have found that the fitted parameters $a$ and $b$ in the generated networks are close to those in the benchmark email network.

However, in the generated networks, the maximum degree is 155, and there are 15 vertices whose degrees are more than 155 in the benchmark network. Moreover, the maximum degree is 345 in the benchmark network. The 15 vertices can be considered as the "special vertices" in the real email network, and the occurrence of those vertices cannot affect the global structure of the network.

Moreover, the Coefficient of Determination (COD) $R^2$ for the generated networks in **Figure 6** is smaller than that in **Figure 4**. This is probably due to that the values of $\sigma$ in the generated email networks are larger than those in the generated BA scale-free networks. In the proposed paradigm, we use the function `std::normal_distribution<double> distribution(0, σ )` to create random numbers, and the sum of the probabilities for creating the random numbers in the range of -5 to 5 (excluding 5) is approximately larger than 0.95. Then, we scale and transform the selected range as follows: `numVert / 2 + (int) numVert / 10 * randomNumber`, where the `numVert` is the number of vertices in the expected network.

The selected range is first scaled to be -0.5 ~ 0.5, then enlarged to be -0.5 * numVert ~ 0.5 * numVert, and finally transformed to be 0 ~ numVert with the offset of 0.5 * numVert.

After scaling the selected range from -5 ~ 5 to -0.5 ~ 0.5, we generate a series of random numbers that conforming to the Gaussian distribution. We have found that more than 95% of the generated random numbers fall into the range of -0.5 ~ 0.5; and there also have very small possibility for creating random numbers that are out of the range of -0.5 ~ 0.5.

With the increase of $\sigma$, the possibility for creating random numbers that are out of the range of -0.5 ~ 0.5 increases. If those random numbers that are out of the range of -0.5 ~ 0.5 are generated, we will ignore and not use them as the indices of vertices. Therefore, with the increase of $\sigma$, the $R^2$, in general, would decrease, i.e., the Goodness of Fit would get worse.

Moreover, it has been observed that the combinations P1D1, P2D1, P0D3, P1D3, and P2D3 can be employed to create complex networks with Power-low degree distributions when the numbers of vertices and numbers of edges in the four benchmark BA scale-free networks are given. However, when employing the above five combinations and when giving the number of vertices and the number of edges in the benchmark email network, it is not able to create complex networks with Power-low degree distributions; instead, those complex networks with Poisson distributions would be created.

It has been observed that all of the above 5 combinations involving the probabilities conforming to the Uniform distribution and Gaussian distribution for selecting vertices. The probabilities conforming the Uniform distribution are uniform. And most of the probabilities conforming the Gaussian distribution are close to the expectation of the Gaussian distribution. For the combination of the Uniform distribution and Gaussian distribution, the discrete of Gaussian distribution would increase with the influence of the Uniform distribution on the probabilities. And this can be deemed as a competition between the Uniform distribution and Gaussian distribution.

And the competition between the Uniform distribution and Gaussian distribution impact the degree distributions of the expected networks. The number of edges is much larger than the number of vertices in the real email network, in this case, the Uniform distribution is of an advantage over the competition between the Uniform distribution and Gaussian distribution. And the degree distributions in combinations is the Poisson distribution with the possibilities of selecting vertices which conforming the Uniform distribution, such as the degree distribution in the combination P0D1.

However, the numbers of vertices are close to the numbers of edges in the benchmark BA scale-free networks, see **Table 2**. In this case, the Gaussian distribution is of an advantage over the competition between the Uniform distribution and the Gaussian distribution. And the degree distributions in combinations is the Power-low distribution with the probabilities of selecting vertices which conforming the Gaussian distribution, such as the degree distribution in the combination P0D2.

## 4. Discussion

In this section, we will analyze the advantages and shortcomings of the proposed paradigm, and point out our future work.

### 4.1 Advantages of the Proposed Paradigm

Our paradigm is straightforward. The essential idea behind the proposed paradigm is to select two vertices of each edge according to probability distributions. Different combinations of the probability distributions can lead to different degree distributions in generated networks. Moreover, compared with other relevant approaches for generating complex networks, our paradigm is of low computational complexity. The computational complexity of both the first and the second procedures are nearly $O(n)$, while the computational complexity of the third procedure is $O(nlogn)$.

Our paradigm is generic. Different combinations of probability distributions for selecting vertices can lead to different complex networks, such as the BA scale-free network, the WS small-world network, scale-free networks with different power exponents, or other complex networks when using other combinations of probability distributions for selecting vertices.

Our paradigm is of strong applicability. The main difference between our paradigm and other relevant approaches is that: besides the classic BA network and the WS network, the proposed paradigm can also be employed to generate complex networks with expected numbers of vertices, numbers of edges, and types of degree distributions, if appropriate probability distributions for selecting vertices are employed.

### 4.2 Shortcomings of the Proposed Paradigm

In the proposed paradigm, it is needed to conduct several tests to determine the optimal parameters of the probability density function. If a real complex network is expected, the optimal parameter of selecting probability distribution needs to be found by tests. In our paradigm, two vertices of each edge are selected according to specific probability distributions. The parameters of the probability density functions are of strong effect on the degree distributions of the generated networks. Therefore, several tests need to be conducted to determine optimal parameters of the probability density functions. For example, we generate the real email scale-free network by adjusting the variance $\sigma$ of the Gaussian distribution, and the optimal $\sigma$ is 1.4, 1.45, and 1.4 in the combination P0D2, P1D2, and P2D2, respectively.

Currently, our paradigm cannot be used to generated those networks including several vertices with extremely large degrees. However, the occurrence of those vertices with extremely large degrees is of weak influence on the global network structure. For example, there are 15 vertices with extremely large degrees in the benchmark real email network, but there are no such vertices in the generated email network.

### 4.3 Future Work

In this paper, we generate the scale-free networks and small-world networks by using two probability distributions in our paradigm, i.e., the Uniform distribution and Gaussian distribution. Other complex networks can also be generated by using several other appropriate probability distributions. This means our paradigm can be employed and applied to generate unknown network structures.

## 5 Conclusion

In this paper, we have proposed a simple and generic paradigm for creating various complex networks using the strategy of vertex selecting-and-pairing. The essential idea behind the proposed paradigm is straightforward: first two vertices are randomly selected from a given set of vertices and then paired into an edge. The expected complex network could be generated after forming a given number of edges. The different choices of random selection could be used to create different types of complex networks, including the classical small-world networks and scale-free networks. To demonstrate the effectiveness of the proposed paradigm, several created and real-world complex networks have been used for validation. The most obvious inherent feature of the proposed paradigm is the simplicity and generality, which indicate that the proposed paradigm could be easily and widely used in dealing with the problems in various complex systems.

**Acknowledgments** This work was supported by the Natural Science Foundation of China (Grant Numbers 11602235).